# Benchmarking the True Random Number Generator of TPM Chips

Alin Suciu
*Technical University of Cluj-Napoca*
*Alin.Suciu@cs.utcluj.ro*

Tudor Carean
*Technical University of Cluj-Napoca*
*djpitagora@yahoo.com*

## Abstract

*A TPM (trusted platform module) is a chip present mostly on newer motherboards, and its primary function is to create, store and work with cryptographic keys. This dedicated chip can serve to authenticate other devices or to protect encryption keys used by various software applications. Among other features, it comes with a True Random Number Generator (TRNG) that can be used for cryptographic purposes. This random number generator consists of a state machine that mixes unpredictable data with the output of a one way hash function. According the specification it can be a good source of unpredictable random numbers even without having to require a genuine source of hardware entropy. However the specification recommends collecting entropy from any internal sources available such as clock jitter or thermal noise in the chip itself, a feature that was implemented by most manufacturers. This paper will benchmark the random number generator of several TPM chips from two perspectives: the quality of the random bit sequences generated, as well as the output bit rate.*

## 1. Introduction

Trusted Computer Group was created in 2003 by the collaboration of several well known private companies like Microsoft, IBM, Intel, and HP, with the purpose of producing a set of specifications for a "trusted hardware platform" and for a chip called Trusted Platform Module (TPM).

These specifications include both hardware recommendations as well as API specifications (known as TSS – Trusted Software Stack) to be used by software programs interfacing with the chip. The first set of specifications released, 1.1, had very little success on the market. The current version, released in 2007 is 1.2, and was already implemented by several different hardware manufacturers. It is expected to become an ISO standard by the end of this year.

The TPM is a small microchip that comes with some of the new motherboard models. Its main purpose is to generate, store and protect cryptographic keys. The fact that it's a hardware component allows it to authenticate other hardware devices and makes it less vulnerable to software attacks. Obviously the chip can only store a small limited amount of encryption keys in its special registers called platform configuration registers (PCRs), so the idea is to hold a master key that is used to encrypt all the other keys.

Among many other things the TPM chip also contains a true random number generator designed to be used for cryptographic purposes. This paper will benchmark the random number generator of several such chips, made by different manufacturers from two perspectives: the quality of random numbers, as well as the output rate. For this purpose a Windows application was developed to interface with the chip and gather numbers for testing. We will also demonstrate that even though all chips adhere to the same standard imposed by TCG they operate in different ways and most likely contain different components.

## 2. State of the art in the use of TPM

Even though the TPM 1.1 standard already exists for a few years it never became too popular because of lack of built-in support from the most popular operating systems, incompatibilities between drivers produced by different manufacturers, as well as the fact that the chip itself is only shipped with some new motherboards, mostly on laptops. Even though many desktop motherboards support it, the module needs to be ordered separately. The more recent standard however, TPM 1.2 slowly gained popularity because of the native support for TPMs in Windows Vista, as well it's usage in the new hard disk security solution BitLocker from Microsoft.

The TPM support in Windows Vista comes in the form of Trusted Base Services (TBS), a component meant to work with any TPM and to replace the manufacturer produced low level driver (TDDL - Trusted Device driver library). This would enable an application to interface with a TPM regardless of brand, and avoid issues like the different naming schemes for DLL files containing the relevant APIs, or using a different SDK for each brand.

A similar solution concerning the standardization of the APIs through a common open source TSS is provided on Linux by TrouSerS (The open-source TCG Software Stack) and in java by TPM/J library. Unfortunately both require drivers underneath. The latter is using TBS when running under Vista, but under any other Windows operating system it requires an Infineon driver which is obviously only available if the TPM is of the same brand.

Even though all TPMs follow a standard, much of their internal construction is intentionally left a black box by the documentation, not only giving manufacturers a free hand on how to implement the requirements but also to protect the security of the device by keeping vendor specific functionality a secret. Since the chip needs to perform sequential operations either a CPU or a microcontroller is required. Different chips may have completely different implementation. It is safe to say that the key requirements needed include a CPU or microcontroller, non-volatile and volatile memory, and input/output (I/O) communication support. Some chips may include other components than these bare minimums, such as co-processors or security circuits for protecting the chips from different kind of attacks.

## 3. Benchmarking the random number generator of the TPM chips

In order benchmark the speed and test the quality on the random numbers generated by the TPM, we created a program in Visual C++, using only the standard libraries that come with Windows Vista (Trusted Base Services) and the Windows Platform SDK. The only requirement is of course that the TPM be enabled in BIOS and activated.

This solution was chosen in favor of using a manufacturer supplied TSP layer (Trusted Service Provider) of the Trusted Software Stack, to ensure full compatibility with any brand of chips. The latter solution is more high level, meaning that the programmer only needs to call high level functions such as Tspi_TPM_GetRandom() to accomplish the desired tasks, as described in the TSS Specifications issued by TCG. However the implementation of these function not only reside in differently named dynamic link libraries for each brand, but are sometimes called differently. For example while Sinosun's libraries export these functions directly, only needing the header files supplied by TCG to use them, Infineon's libraries export classes using the COM interface model and require the use of their SDK.

The TBS that comes with Windows Vista only offers low level functions such Tbsip_Submit_Command(), functions that are analog with those in those from the manufacturer produced low level driver (TDDL). This one for instance represents the means of calling any other feature of the TPM by accepting as argument a buffer containing commands in a format specified by the „TPM Commands Specification Document" issued by TCG. The output is another buffer of data, whose structure is specified by the same document.

For instance using the random number generator must be done by sending the raw command in table 1. The output comes as a buffer structured as described in table 2.

**Table 1. TPM_GetRandom input message**

| Size (bytes) | Field | Description |
|---|---|---|
| 2 | Authorization Tag | No authorization is necessary for this command (0xC1) |
| 4 | Parameter size | The size of request header (0x0C) |
| 4 | Ordinal | Function code (0x46) |
| 4 | Bytes Requested | An integer that decides how many numbers will be returned |

**Table 2. Output message for TPM_GetRandom**

| Size (bytes) | Field | Description |
|---|---|---|
| 2 | Authorization Tag | |
| 4 | Parameter size | Size of the response |
| 4 | Return code | 0 means success, failure otherwise. |
| 4 | Random bytes size | The size of the random number block returned |
| depends | Random numbers | |

The application developed to assist in achieving the goals of this paper presents the user with the interface from figure 1.

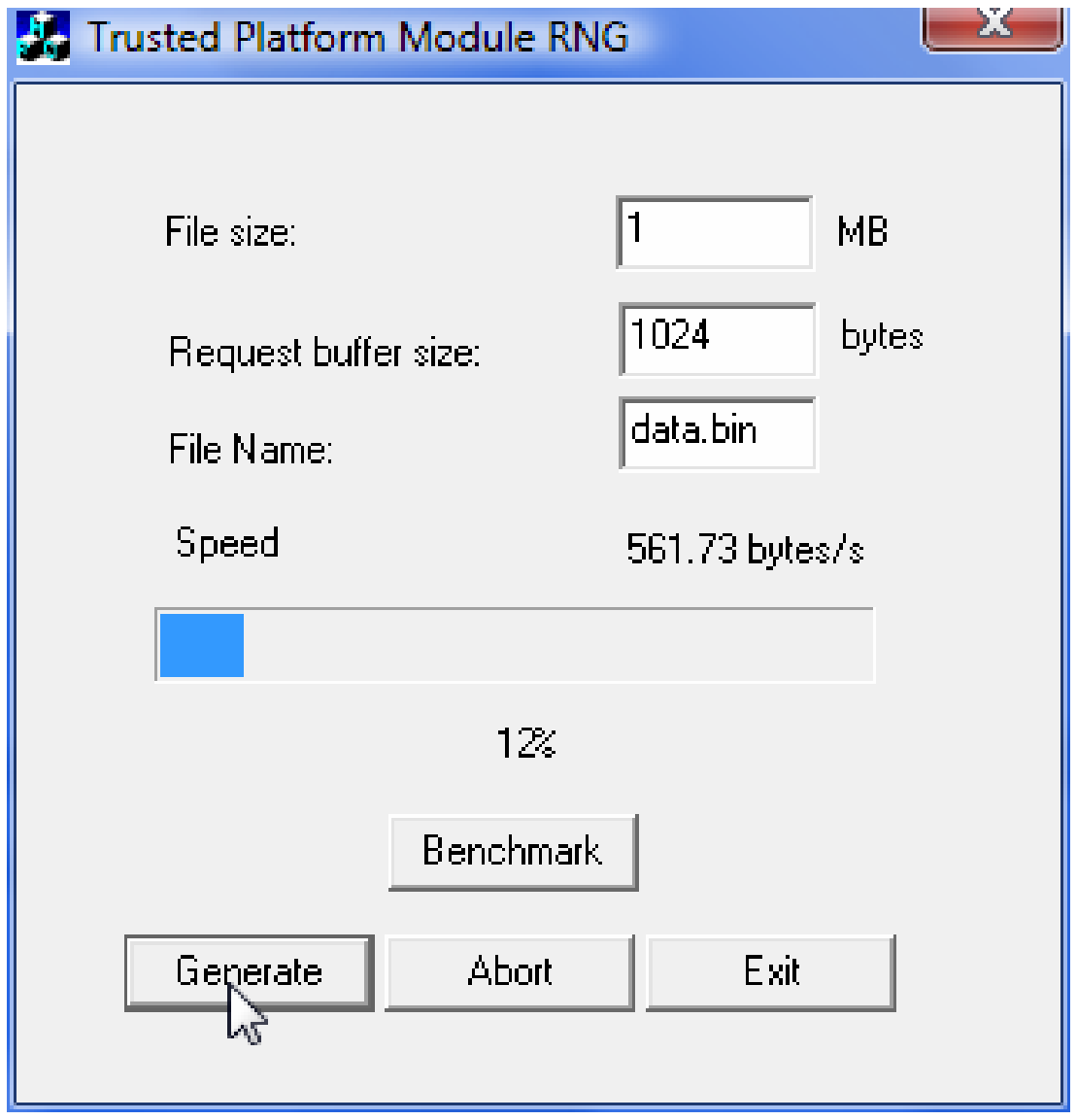

**Fig 1. Application interface**

The user can enter a filename where the generated data is to be stored and the desired amount of data. The “request buffer size” option allows the user to change the number of bytes of random data collected by one call to the TPM. Pressing the button “Generate” starts a background worker thread that collects the numbers from the chip. The time required for each call to the TPM is measured and summed in order to show the user an average speed that is updated once in a while. If for whatever reason the user decides to cancel the process he can do that by pressing the “Abort” button.

The GetRandom function accepts as buffer size any number of bytes between 1 and 2048. Obviously getting 1 byte of random data is not the same in terms of speed as getting 2048, so one of the features of the developed application is to benchmark the time required for the TPM to respond to requests of different sizes. This feature of the program is accessed by using the Benchmark function. The test consists of a loop that varies the requested number of bytes from 1 to 2048 and writes down in a CSV file the time needed for the call to complete. The CSV format was chosen because it can easily be imported into Excel, where one could better study the benchmark data and put it in a graphic representation

## 4. Experimental Results

We have done the test by running the application described above on four machines (two desktop computers and two laptops) each with a different brand of TPM 1.2 chip: Infineon, Atmel, Intel and Sinosun.

To insure compatibility between the different chips, as well as the same conditions in term of software and drivers, all tests were performed on Windows Vista using its native TBS driver.

### 4.1. Output speed tests

The idea behind these tests is to find out how fast each TPM responds to a query for random numbers of a certain size. The TCG standards specify that TPMs should be able to return random numbers in batches of 1 to 2048 bytes. However 3 of the 4 TPMs tested had a much lower maximum request sizes. This value is 1259 bytes for Infineon chips, 1226 bytes for Intel, 768 for Atmel and 2048 for Sinosun.

Making a larger request doesn’t result in an error, but the returned headers clearly indicate a different buffer output size. An application that doesn’t check these headers to validate that its request was fully executed could mistakenly use numbers that are not random, or were already in the buffer prior to making the TPM call.

According to the tests, the efficiency of the generator depends on how many numbers it is requested to generate. To insure the results are not influenced by external factors such as other OS processes each request was repeated 10 times and the only average needed time was exported to Excel and was plotted on the graphics.

Figure 2 shows how the Infineon chip clearly has two internal operation modes, each with a linear speed increase. The second line on the graphic is most probably the result of a condition that adds an extra operation from time to time, such as reseeding. One in every few dozen requests is on this second line.

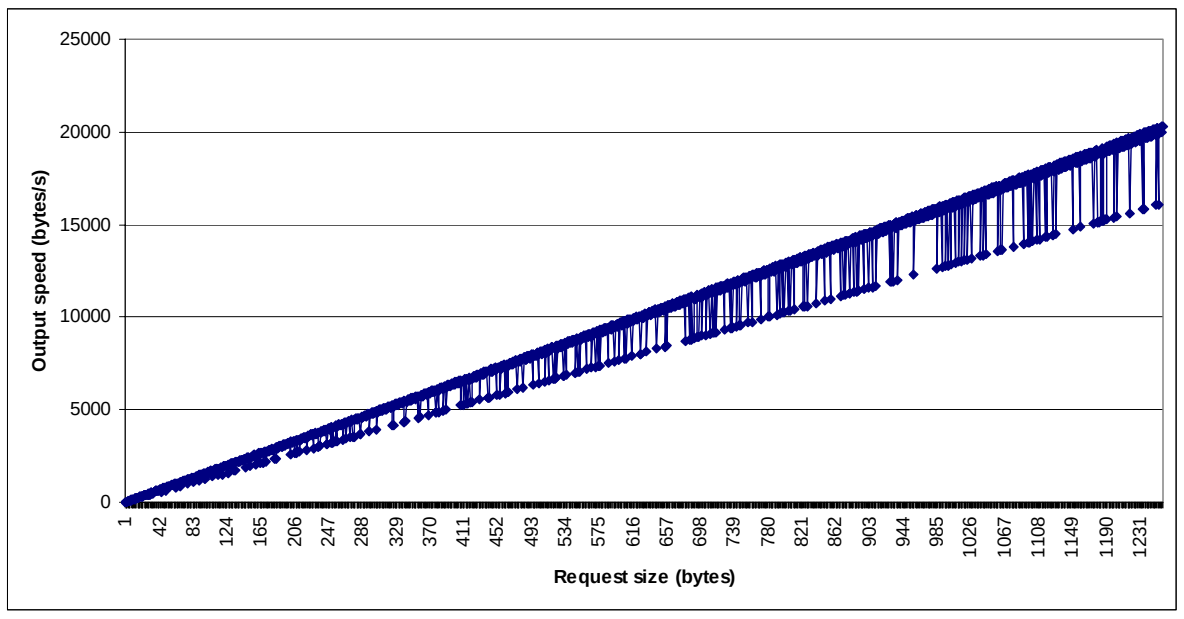

**Fig 2. Infineon TPM speed test**

Figure 3 shows that Intel TPM’s output speed quickly increases between 1 and 400 bytes request size, and remains relatively constant (in the range of 500 bytes/s) for higher request sizes.

What is really interesting about this chip is that not only the operation isn’t linearly dependent on the “workload” as expected, but the graphic has discontinuities approximately 64 bytes apart, which indicates a totally different algorithm/method of obtaining the random numbers. A possible explanation of these discontinuities could be a continuous gathering of random numbers in a fixed size buffer, and serving of the requests from it.

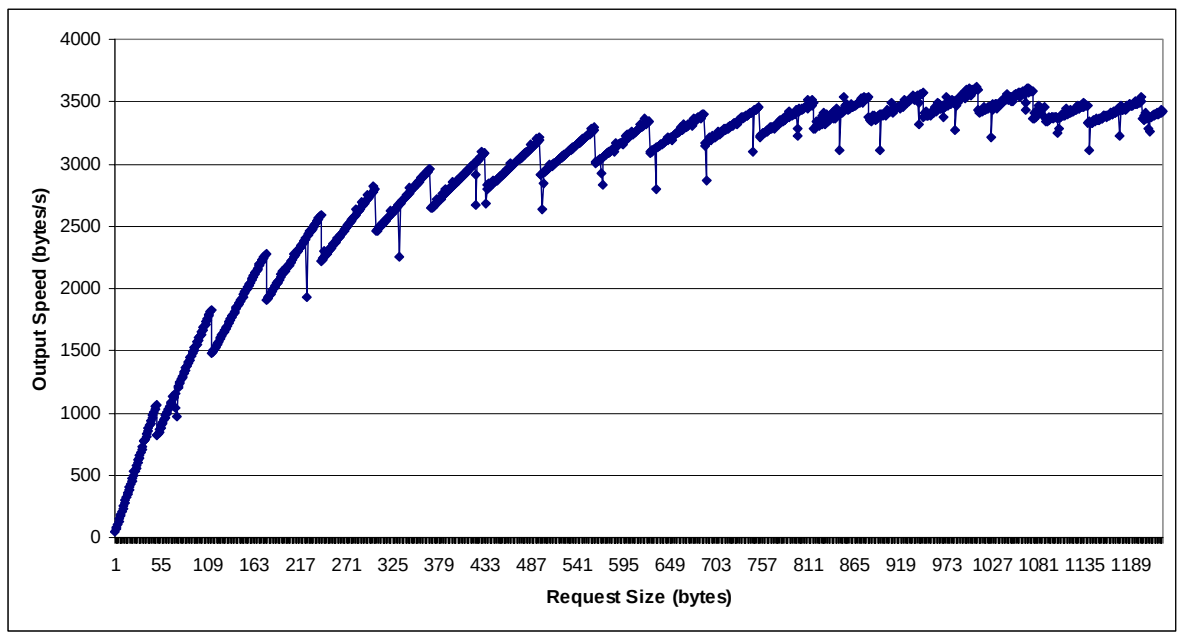

**Fig 3. Intel TPM speed test**

Atmel chips on the other hand use a linear algorithm that unfortunately doesn’t scale quite as well as Infineon’s. The graphic shows a discontinuity at 538 bytes, where the output speed drops significantly, then linearly grows again at the same rate. This behavior suggests that any request higher then 538 bytes is actually handled in two similar steps by the chip. The reason behind this could be an internal buffer of size around that value.

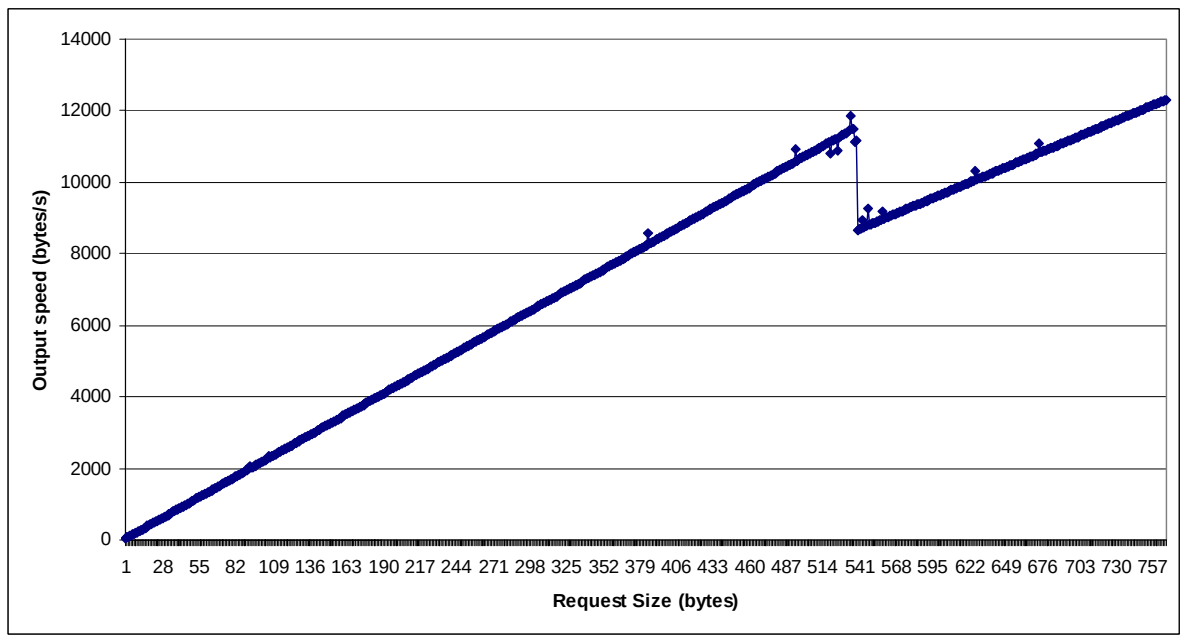

**Fig 4. Atmel TPM speed test**

Out of the chips tested the Sinosun TPM is the only one that respects the specifications to the letter concerning the maximum request size allowed, however it also gave the lowest output speed, which is almost 30 times smaller then that of the fastest chip (Infineon). If zoomed, the graphic in figure 5 looks very similar to the one for Atmel chips.

The output speed rises fast up until 300 bytes request size and remains almost constant (plus minus 20 bytes/sec) for any value higher then that. The same kind of discontinuities can be observed, this time around 20 bytes apart. Repeated experiments show the discontinuities are in the same places every time. A reasonable conclusion would be that somewhere in the chip there is a 20 bytes buffer where these numbers are gathered.

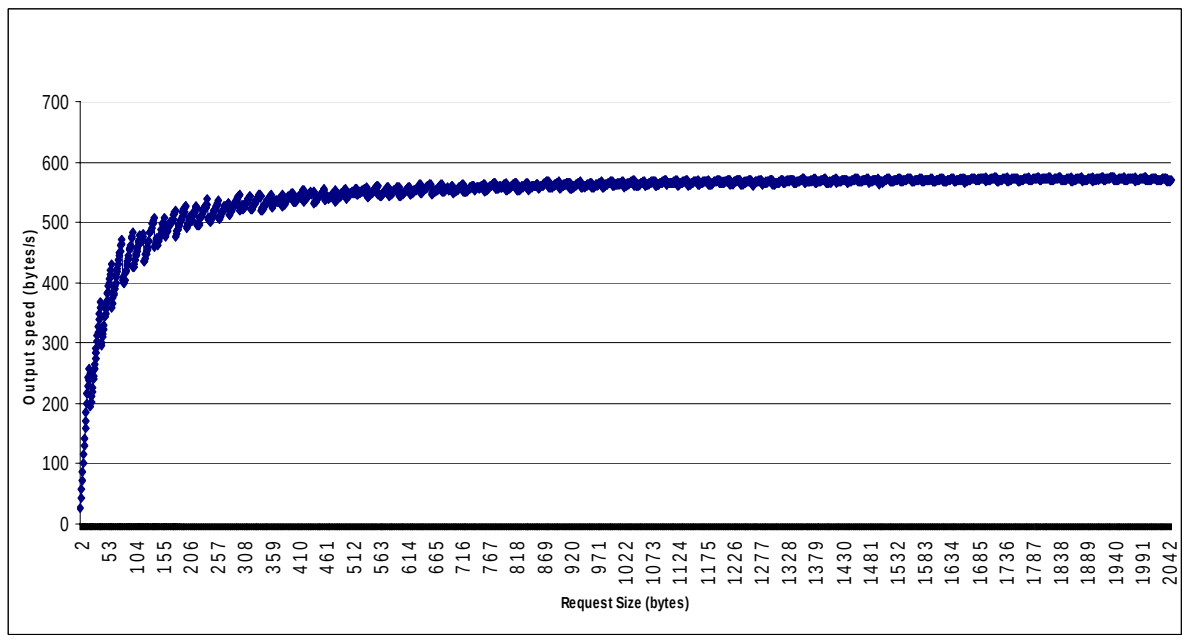

**Fig 5. Sinosun TPM speed test**

## 4.2. Quality tests

The quality of the resulted random numbers was tested with NIST, RNGmeter, Ent, and a Windows port of TestU01. A total of 100 MB of random data was gathered from each TPM and analyzed both as a whole, as well as broken in 10 pieces of 10 MB each.

A first comparison between the generators can be seen in table 3, which contains a summary of the Ent tests on the whole 100 MB files. The first thing that needs to be noticed is that the Intel TPM chip has failed the chi square distribution test on a bit level and is at the suspicion border on a byte level test (10% chi square).

The Sinosun chip is also suspicious according to the same test at bit level with a 5% chi square value that even though is not a bad failure, it casts a shadow of doubt on the randomness of the sequence. Ironically these two chips also have the best result on the serial correlation test, both on byte and bit level.

According to Ent’s algorithms the value it calculates is supposed to be as close to 0 as possible on a random sequence. Also using numbers from the Intel chip the pi value calculated by a Monte Carlo algorithm obtained the smallest error, 0.00%.

Although a few suspicious chi square values were found, all four 100 MB sequences obtain the same score in RNGmeter: 27.6+, indicating no major failures in any of the generators.

**Table 3. The comparison of test results from Ent**

| | Infineon | Intel | Atmel | Sinosun |
|---|---|---|---|---|
| byte level | | | | |
| entropy | 7.99 | 7.99 | 7.99 | 7.99 |
| chi square | 50% | 10% | 50% | 25% |
| Arithmetic mean | 127.4917 | 127.5118 | 127.4991 | 127.4898 |
| Monte Carlo pi | 0.03% | 0.00% | 0.01% | 0.01% |
| Serial correlation | 0.000085 | 0.000132 | 0.000037 | -0.000002 |
| bit level | | | | |
| entropy | 1 | 1 | 1 | 1 |
| chi square | 75% | 10% | 50% | 25% |
| Arithmetic mean | 0.5000 | 0.5000 | 0.5000 | 0.5000 |
| Monte Carlo pi | 0.03% | 0.00% | 0.01% | 0.01% |
| Serial correlation | 0.000019 | 0.000012 | 0.000015 | 0.000003 |

The four 100 MB sequences were also tested with two batteries from TestU01: Rabbit and Alphabit. The Rabbit battery consists of 40 tests out of which the Intel TPM has failed one (Close Pairs Bit Match, t=4) and the Atmel TPM another one (Multinomial BitsOver). The Infineon's random number generator is the only one to fail one of Alphabit's 16 tests: a random walk for L=64. The sequence from the Sinosun chip is the only one to pass all tests from these batteries.

The proportion of failed test on smaller 10 MB files was the same. One of the 10 sequences from the Infineon chip failed two tests of the Rabbit battery: MultinomialBitsOver, and the same random walk it failed for the 100 MB file. Four sequences from the Intel chip each failed a test: PeriodsInString, ClosePairsBitMatch, Run of bits and MultinomialBitsOver.

Two sequences from the Atmel TPM also failed a test each: HammingIndep and HammingCorr. Despite the great results on the large 100 MB file, the MaultinomialBitsOver test failed on one of the Sinosun TPM's 10 MB file.

Only three out of the NIST tests were not performed: runs, DFT and Universal statistical test. When testing the big 100 MB sequences all four generators failed the same test only: overlapping templates. However when testing each 10 MB piece individually there was generally more then one failure found in each of them. This might be explained by the fact that the more numbers there are in the sequence the more they compensate for the lack of quality in some of the blocks.

Table 4 shows how many of the 10 sequences tested for each chip has failed a certain test. The number in the parentheses shows how many tests of the same kind were failed. There aren't any big differences in quality but one can immediately notice that the Infineon and Atmel chips performed better. This doesn't come as a big surprise since the results from Ent and TestU01 already implied it.

**Table 4. Number of sequences that failed each NIST test**

| Test/TPM | Infineon | Intel | Atmel | Sinosun |
|---|---|---|---|---|
| non-periodic template (1) | 2 | 5 | 3 | 4 |
| non-periodic template (2) | 2 | - | 3 | 1 |
| non-periodic template (3) | 2 | 1 | 1 | 2 |
| non-periodic template (4) | - | - | 1 | 1 |
| overlapping template | 2 | 3 | 3 | 3 |
| Serial (both) | 1 | - | - | - |
| Block frequency | - | 1 | - | - |
| Frequency | - | 1 | - | - |
| Random excursions | 1 | 2 | 2 | - |
| Random excursions (3) | - | - | - | 1 |
| Random excursions variant (7) | - | - | - | 1 |
| Rank | - | - | - | 1 |
| Cumulative sums (forward and reverse) | - | 1 | - | - |

All generators had one sequence that passed all tests, except the one from the Infineon chip that had three. The most common failed tests are overlapping templates and non-periodic templates. Most of the sequences failed from 1 to 3 non-periodic template tests out of the 149 done by NIST on each 10 MB file. One 10 MB sequence in particular taken from the Sinosun TPM has performed very poorly in comparison to the rest (including from the same chip), failing one non-periodic template test, 3 rand excursions, 7 random excursion variant, overlapping templates and rank test.

Since the Sinosun chip has the slowest generator some of these numbers were not generated in the same day, so the sequence in question may have been generated under totally different conditions than the rest. The Intel TPM also had two sequences (generated in the same run this time) that failed quite of few tests: block frequency, frequency overlapping template, both cumulative sums and random excursions.

## 5. Conclusions

The random number generators tested behaved very different, especially in terms of speed where the fastest chip generated numbers at a rate 30 times higher then the slowest, but also in quality of the random data.

As the output speed graphics suggest, the chips contain different types of components and use different methods of generating random numbers, even though they are bound by the same specification and standard.

While the quality of the output is generally great, the speed by which the numbers were generated on the test system suggests this method of gathering random numbers should rather be used for seeding other generators than for a Monte Carlo simulation.

The TPM remains though a great tool for what it was designed in the first place: storing encryption keys and authenticating other hardware.